# Investigating the Potential of Large Language Model-Based Router Multi-Agent Architectures for Foundation Design Automation: A Task Classification and Expert Selection Study

Sompote Youwai[*], David Phim, Vianne Gayl Murcia, Rianne Clair Onas
AI Research Group, Department of Civil Engineering, King Mongkut's University of Technology Thonburi
[*]Corresponding author

Abstract

This study investigates router-based multi-agent systems for automating foundation design calculations through intelligent task classification and expert selection. Three approaches were evaluated: single-agent processing, multi-agent designer-checker architecture, and router-based expert selection. Performance assessment utilized baseline models including DeepSeek R1, ChatGPT 4 Turbo, Grok 3, and Gemini 2.5 Pro across shallow foundation and pile design scenarios. The router-based configuration achieved performance scores of 95.00% for shallow foundations and 90.63% for pile design, representing improvements of 8.75 and 3.13 percentage points over standalone Grok 3 performance respectively. The system outperformed conventional agentic workflows by 10.0 to 43.75 percentage points. Grok 3 demonstrated superior standalone performance without external computational tools, indicating advances in direct LLM mathematical reasoning for engineering applications. The dual-tier classification framework successfully distinguished foundation types, enabling appropriate analytical approaches. Results establish router-based multi-agent systems as optimal for foundation design automation while maintaining professional documentation standards. Given safety-critical requirements in civil engineering, continued human oversight remains essential, positioning these systems as advanced computational assistance tools rather than autonomous design replacements in professional practice.

Keywords: Router-based multi-agent systems, Large language models, Foundation design, Geotechnical engineering, Task classification, Expert selection, AI-assisted engineering

## 1. Introduction

The integration of artificial intelligence into engineering design processes has emerged as a transformative paradigm, particularly in structural and geotechnical engineering where computational complexity and safety-critical decision-making converge. Recent advances in large language models (LLMs) and multi-agent workflow systems present unprecedented opportunities for automating routine engineering calculations while maintaining adherence to established design codes and standards. Foundation design represents a cornerstone of structural engineering practice, requiring sophisticated analysis of soil-structure interaction, compliance with rigorous safety standards, and integration of complex geotechnical parameters. Current design methodologies for shallow foundations and pile systems rely heavily on established analytical frameworks that provide systematic approaches for bearing capacity calculations, settlement analysis, and safety factor implementation.

The emergence of router-based multi-agent architectures (Salvador Palau et al., 2019) and intelligent task classification systems has demonstrated remarkable potential for complex technical tasks requiring specialized domain knowledge integration. These systems employ router agents to classify incoming design problems and direct them to specialized expert agents, exhibiting capabilities for automating engineering calculations while maintaining traceability and compliance with established design procedures. Recent developments in state-of-the-art models including DeepSeek R1(DeepSeek-AI, 2025), ChatGPT 4 Turbo (OpenAI, 2023), Grok 3 (xAI, 2025), and Gemini 2.5 Pro (Google DeepMind, 2025) have shown enhanced reasoning capabilities in mathematical computation, technical documentation generation, and engineering problem-solving across diverse domains.

Recent systematic reviews of Large Language Model (LLM) applications in construction engineering have identified seven primary implementation domains: training and education, planning and scheduling, safety analysis and hazard recognition, document generation and compliance checking, specialized virtual assistants, code generation, and Building Information Modeling functionalities (Kampelopoulos et al., 2025). Emerging research has demonstrated computational engineering applications through domain-specific frameworks with varying degrees of success. Liang et al. ((2025) developed an automated structural analysis framework integrating LLMs with finite element analysis software, achieving 100% accuracy using GPT-4o on a benchmark dataset of 20 structural analysis word problems through implementation of domain-specific prompt engineering and in-context learning methodologies. Performance evaluation by Chen et al. (2024) demonstrated accuracy improvements from 28.9% with zero-shot learning to 67% with custom instructional prompting when testing GPT-4 on 391 geotechnical engineering questions. Xu et al. (2025) implemented multi-GeoLLM, a multimodal framework incorporating Retrieval-Augmented Generation (RAG) search engines, self-review modules, mathematical computation tools, and human-in-the-loop feedback mechanisms for geotechnical design applications. These implementations employ monolithic agent architectures that process heterogeneous engineering tasks through singular computational pathways, with RAG-dependent methodologies exhibiting inherent limitations including performance correlation with retrieval architecture quality, semantic drift across engineering terminology, degraded performance on novel problem types, computational inefficiency during real-time operations, and constraint to pre-existing example databases rather than generative problem-solving capabilities.

The absence of routing agent frameworks specifically designed for foundation design applications constitutes a significant methodological gap in current LLM implementations for engineering calculations. Existing approaches demonstrate fundamental limitations through reliance on pattern-matching retrieval systems and continuous human oversight, which constrains system scalability and introduces systematic error propagation across heterogeneous problem domains with variable computational complexity requirements. Current validation methodologies exhibit limited scope with datasets predominantly focused on shallow foundation systems, lacking comprehensive evaluation across pile foundations, combined foundation systems, and complex soil-structure interaction scenarios that represent critical components of contemporary foundation engineering practice. This research addresses these limitations through implementation of a routing agent framework that dynamically classifies problem typologies and directs computational tasks to specialized domain-expert agents operating through tailored system prompts and contextual expertise rather than RAG retrieval mechanisms. The study will conduct systematic benchmarking against state-of-the-art reasoning-based models, tree search architectures, Grok 3, and Gemini Pro 2.5, utilizing current frontier model capabilities as the primary computational engine. The research methodology will expand evaluation protocols beyond conventional accuracy metrics to encompass comprehensive assessment of computational consistency, reliability characterization, and systematic failure mode analysis across the complete spectrum of foundation design applications encountered in professional engineering practice, thereby addressing critical impediments to deployment in safety-critical structural engineering applications requiring autonomous computational capabilities.

The primary objectives of this investigation are: (1) to evaluate the comparative performance of router-based multi-agent workflows against baseline models in foundation design calculations, (2) to assess the effectiveness of designer-checker multi-agent architectures in improving calculation accuracy and error detection, (3) to examine the capability of router-based agent systems in handling complex geotechnical scenarios and specialized calculations, and (4) to identify critical limitations and validation requirements for practical implementation in geotechnical engineering practice.

The significance of this research extends beyond technological advancement to establish empirical evidence regarding the superiority of router-based multi-agent approaches over current state-of-the-art single-model implementations in safety-critical engineering applications. The findings will inform future development of intelligent design assistance tools while establishing realistic expectations for AI capabilities in geotechnical applications, ultimately supporting the responsible advancement of multi-agent AI technology in engineering practice while maintaining the essential human oversight required for public safety and professional accountability in structural design.

This research makes several significant contributions to the fields of artificial intelligence applications in engineering and geotechnical design automation:

- This study introduces the first comprehensive router-based multi-agent system specifically designed for foundation engineering applications, incorporating intelligent task classification and expert selection mechanisms. The proposed architecture demonstrates a systematic approach to automating complex geotechnical calculations through specialized agent coordination, representing a significant advancement over conventional sequential AI workflows in engineering domains. The system employs a dual-tier classification framework that successfully distinguishes between pile and shallow foundation problems, enabling the application of fundamentally different analytical approaches appropriate to each foundation type.
- The research establishes a rigorous evaluation methodology for assessing AI model performance in geotechnical engineering applications, comparing router-based multi-

- agent systems against state-of-the-art baseline models including DeepSeek R1, ChatGPT 4 Turbo, Grok 3, and Gemini 2.5 Pro. This framework provides standardized metrics for accuracy assessment, chain-of-thought reasoning evaluation, complex scenario handling, and output consistency across diverse foundation design tasks. The evaluation protocol encompasses twenty-seven distinct test cases spanning seven primary categories of geotechnical engineering problems, with triple-trial execution to ensure statistical robustness.
- The study provides quantitative evidence demonstrating that router-based multi-agent architectures achieve superior performance compared to individual state-of-the-art models, with exceptional performance scores of 95.00% for shallow foundation design and 90.63% for pile design, representing improvements of 8.75 and 3.13 percentage points over standalone Grok 3 performance respectively. The router-based approach consistently exceeded conventional workflow performance by margins ranging from 10.0 to 43.75 percentage points across different foundational models.
- The study establishes practical considerations for deploying AI-assisted foundation design systems in professional engineering practice, including validation requirements, human oversight protocols, and safety-critical implementation frameworks. The research emphasizes that while router-based systems demonstrate promising capabilities, the critical nature of civil engineering applications necessitates maintaining human oversight and verification protocols due to fundamental public safety concerns inherent in structural design.
- The research demonstrates effective prompt engineering techniques for geotechnical applications, incorporating chain-of-thought reasoning, one-shot learning examples, and domain-specific knowledge integration. The implementation of structured prompt engineering approaches incorporated geotechnical domain knowledge, mathematical formulations, and validation criteria within agent architectures, enabling consistent application of established design procedures while maintaining solution transparency and verification capabilities.
- The study identifies specific areas requiring further investigation, including the development of specialized training datasets for complex geotechnical calculations, uncertainty quantification frameworks for safety-critical applications, and hybrid architectures combining strengths of different AI models. These findings provide a roadmap for future research in AI-assisted engineering design, with particular emphasis on developing domain-specific fine-tuning approaches that incorporate established geotechnical methodologies.
- This research establishes baseline accuracy metrics for AI performance in foundation engineering tasks, providing reference standards for future developments in the field. The comprehensive analysis provides empirical evidence for the varying effectiveness of different AI approaches across specialized domains, establishing critical performance baselines that inform realistic expectations for AI capabilities in geotechnical applications and support the responsible advancement of multi-agent AI technology in engineering practice.

The remainder of this study is organized as follows: Section 2: Related Works provides a comprehensive review of existing literature on artificial intelligence applications in structural engineering, multi-agent systems in engineering design, and current methodologies for foundation design automation, examining previous research in AI-assisted engineering calculations and establishing the theoretical foundation for router-based multi-agent architectures. Section 3: Model Architecture presents the detailed design and implementation of the proposed router-based multi-agent system, including the mathematical formulation of agent components, the dual-tier classification framework, and the integration of specialized tools through LangChain interfaces. Section 4: Experiment outlines the comprehensive experimental methodology employed to evaluate the proposed system, including the evaluation rubric design, test case development, and systematic testing protocol across twenty-seven distinct geotechnical engineering scenarios with triple-trial execution for statistical robustness. Section 5: Results of Model Evaluation presents the detailed experimental findings, including quantitative performance comparisons across different AI model configurations and comprehensive performance tables demonstrating the superior performance of router-based multi-agent systems compared to baseline models. Section 6: Discussion analyzes the implications of the experimental results, explores the mechanisms underlying the superior performance of router-based systems, and addresses the limitations and practical considerations for implementation in professional engineering practice. Section 7: Conclusions summarizes the key findings of the study, reaffirms the contributions to the field of AI-assisted geotechnical engineering, and consolidates the evidence supporting router-based multi-agent systems as a significant advancement in foundation design automation while emphasizing the continued importance of professional engineering oversight.

## 2. Related works

Large Language Models (LLMs) have demonstrated increasing adoption in construction engineering applications, yet systematic investigation of their utilization in engineering calculations remains limited. Kampelopoulos et al. (2025) conducted a comprehensive literature review analyzing previous studies published between 2021-2024, identifying seven primary application domains: training and education (Uddin et al., 2023) , planning and scheduling (Guan et al., 2023), safety analysis and hazard recognition (Smetana et al., 2024), document generation and compliance checking(Pu et al., 2024), specialized virtual assistants (Yang et al., 2024), code and data generation, and Building Information Modeling (BIM)/Building Energy Modeling (BEM) functionalities (Han et al., 2025; Jang and Lee, 2024). Notably, their analysis revealed a complete absence of research attempting direct deployment of LLMs for fundamental engineering calculations, including structural analysis, load computations, and design calculations that constitute the core of construction engineering practice.

This computational gap represents a significant limitation in current LLM implementation within the construction domain. While existing studies demonstrate efficacy in document processing, scheduling optimization, and compliance verification tasks, the lack of investigation into computational engineering applications indicates both methodological constraints and research opportunities. Kampelopoulos et al. (2025) identified inherent accuracy limitations in LLMs when processing numerical data and executing mathematical computations, potentially explaining the observed avoidance of safety-critical engineering calculations where precision requirements are stringent.

However, emerging research has begun addressing LLM applications in engineering calculations. Liang et al. (2025) developed an automated structural analysis framework integrating LLMs with finite element analysis software, where LLMs parse textual structural descriptions and generate executable Python scripts for analysis using OpenSeesPy. The framework implements domain-specific prompt engineering and in-context learning methodologies to enhance problem-solving capabilities, enabling automated analysis from textual input to computational outputs. Performance evaluation using a benchmark dataset of 20 structural analysis word problems (SAWPs) with validated ground-truth solutions demonstrated that GPT-4o achieved 100% accuracy, outperforming GPT-4 (85%), Gemini 1.5 Pro (80%), and Llama-3.3 (30%). Domain-specific instruction implementation yielded 30% performance enhancement for asymmetrical configuration problems, substantially increasing automation levels relative to conventional methodologies. Similarly, Joffe et al. (2025) investigated code compliance verification in civil and structural engineering design, addressing the complexity and time-intensive nature of verifying design compliance with regulatory provisions. Their research presents a framework for developing open-source, scalable LLM-based applications enabling engineers to obtain accurate responses to codes-and-standards queries with corresponding citations through natural language interfaces. Preliminary implementation utilizing the National Building Code of Canada 2020 yielded promising results, demonstrating the framework's potential for assisting design engineers in efficient work completion.

Chen et al. (2024) investigates the utility of large language models (LLMs), specifically GPT-4, in geotechnical engineering education and problem-solving. The researchers evaluated GPT-4's performance using 391 questions from a standard geotechnical engineering textbook, testing three prompting strategies: zero-shot learning, chain-of-thought (CoT) prompting, and custom instructional prompting across various topics and cognitive complexity levels based on Bloom's taxonomy. The study found that while GPT-4 achieved only 28.9% accuracy with zero-shot learning and 34% with CoT prompting, custom instructional prompting significantly improved performance to 67% accuracy by specifically addressing identified error types including conceptual, grounding, calculation, and deficiency errors (particularly related to visual interpretation). The research reveals that GPT-4 performs better on text-based questions compared to image-based ones and shows varying effectiveness across different geotechnical topics and cognitive complexity levels. The authors conclude that while GPT-4 demonstrates potential as an educational tool in geotechnical engineering, particularly for generating questions and providing tutoring support, it requires human oversight and verification, especially for complex problems, and emphasize the need for balanced integration of AI in education alongside continued human expertise.

Currently, Xu et al.(2025) presents multi-GeoLLM , an innovative multimodal framework that integrates multiple large language model agents for intelligent geotechnical design, addressing key limitations in existing LLM applications by incorporating four core modules: a RAG-based search engine that uses Vector Space Modelling to retrieve relevant examples from a pre-built multimodal database and generate optimal geo-prompts, a self-review module with three specialized LLM agents for extracting and validating design information from text and image inputs, a mathematical tool module for precise calculations and logical decision-making, and a Human-in-the-Loop Feedback module for engineer verification and iterative refinement. The RAG implementation works by embedding new design assignments as vectors using transformer models for text and CNNs for images, computing cosine similarity to retrieve the most relevant examples from the database, and then constructing geo-prompts by integrating the assignment with retrieved examples using predefined templates. Validated on 60 unreinforced footing design cases across text, image, and combined datasets, the framework demonstrated exceptional performance.

Current LLM applications in construction engineering demonstrate established efficacy in document processing, scheduling, and compliance verification; however, systematic investigation of computational engineering calculations remains constrained. Kampelopoulos et al. (2025) identified seven primary application domains but revealed no research addressing direct LLM deployment for fundamental engineering calculations. Recent frameworks by Liang et al. (2025) and Xu et al. (2025) have introduced automated structural analysis and geotechnical design through domain-specific prompting and multimodal integration. These implementations employ monolithic agent architectures that process diverse engineering tasks through singular computational pathways, potentially constraining scalability and introducing systematic errors across heterogeneous problem domains with variable complexity requirements. Xu et al. (2025) achieved efficiency through RAG-based vector database systems utilizing pattern-matching retrieval of analogous examples, supplemented by human-in-the-loop feedback mechanisms for iterative output refinement through engineer verification. Such RAG-dependent methodologies exhibit multiple inherent limitations that compromise system reliability. Performance correlates directly with retrieval architecture quality and document segmentation algorithms, where insufficient implementation prevents accurate example identification. Additionally, RAG systems suffer from semantic drift when engineering terminology varies across documents, leading to retrieval of contextually irrelevant examples. The systems demonstrate poor performance with novel problem types absent from training corpora, exhibit computational inefficiency during real-time retrieval operations, and face scalability constraints as database size increases exponentially. Furthermore, RAG approaches cannot distinguish between outdated and current engineering standards, potentially retrieving obsolete design codes or superseded calculation methodologies. The dependency on pre-existing examples also limits creative problem-solving capabilities, as the system cannot generate solutions beyond the scope of retrieved precedents, fundamentally constraining innovation in complex engineering scenarios.

The absence of routing agent frameworks for foundation design applications represents a critical methodological gap. Routing agents can dynamically classify problem types and direct computational tasks to specialized domain-expert agents, each operating independently through tailored system prompts and contextual expertise rather than relying on RAG retrieval mechanisms. This approach offers enhanced accuracy through task-specific specialization, improved error mitigation through distributed validation, and scalable architecture accommodating diverse calculation types without dependence on pre-existing example databases or continuous human oversight. Furthermore, existing validation studies demonstrate narrow scope with limited datasets focused predominantly on shallow foundations, lacking systematic evaluation across pile foundations, combined systems, and complex soil-structure interaction scenarios. This study addresses these limitations by implementing comprehensive benchmarking against state-of-the-art reasoning-based models, tree search architectures, Grok 3, and Gemini Pro 2.5, utilizing current frontier model capabilities as the computational engine. Current evaluation protocols emphasize accuracy metrics without comprehensive assessment of consistency, reliability, and failure mode characterization across the full spectrum of foundation design applications encountered in engineering practice, representing a significant impediment to practical deployment in safety-critical structural engineering applications requiring autonomous computational capabilities.

## 3. Model architecture

This study aimed to enable large language models to function as task-specific design agents through a systematic approach to agent specialization. The methodology involves task classification to designate specialized agents with predefined system prompts that incorporate both design procedures and implementation code. This approach addresses the inherent limitations of large language models that are trained for general-purpose applications rather than domain-specific tasks. While fine-tuning represents an alternative approach for model specialization, it requires substantial computational resources and significant financial investment. Training large-scale models with extensive parameter sets necessitates supercomputing infrastructure with high-capacity memory systems, which are not accessible through conventional computing platforms. The proposed agentic workflow with intelligent routing to expert agents via system prompts presents a more viable alternative to fine-tuning within current technological and economic constraints.

The proposed system architecture is constructed upon the LangChain framework (Chase, 2022) and integrates Google's Gemini-2.5-Pro-Preview and xAI's Grok-3 as the primary large language models. Grok-3 represents an advanced reasoning model trained with reinforcement learning at unprecedented scale, demonstrating superior performance across academic benchmarks in mathematics, coding, and scientific reasoning (xAI, 2025). The system implements a sophisticated multi-agent orchestration methodology within an n8n workflow automation environment (n8n GmbH, 2025), wherein individual agents are configured as domain-specific experts in geotechnical engineering applications. Multi-agent systems in geotechnical monitoring enable decentralized approaches to data acquisition, analysis, and decision-making through seamless interplay between agents (Shakshuki and Reid, 2023). This architectural design facilitates automated task distribution and intelligent workflow routing, thereby enabling efficient computational processing of complex geotechnical analysis scenarios (Herrera et al., 2020).

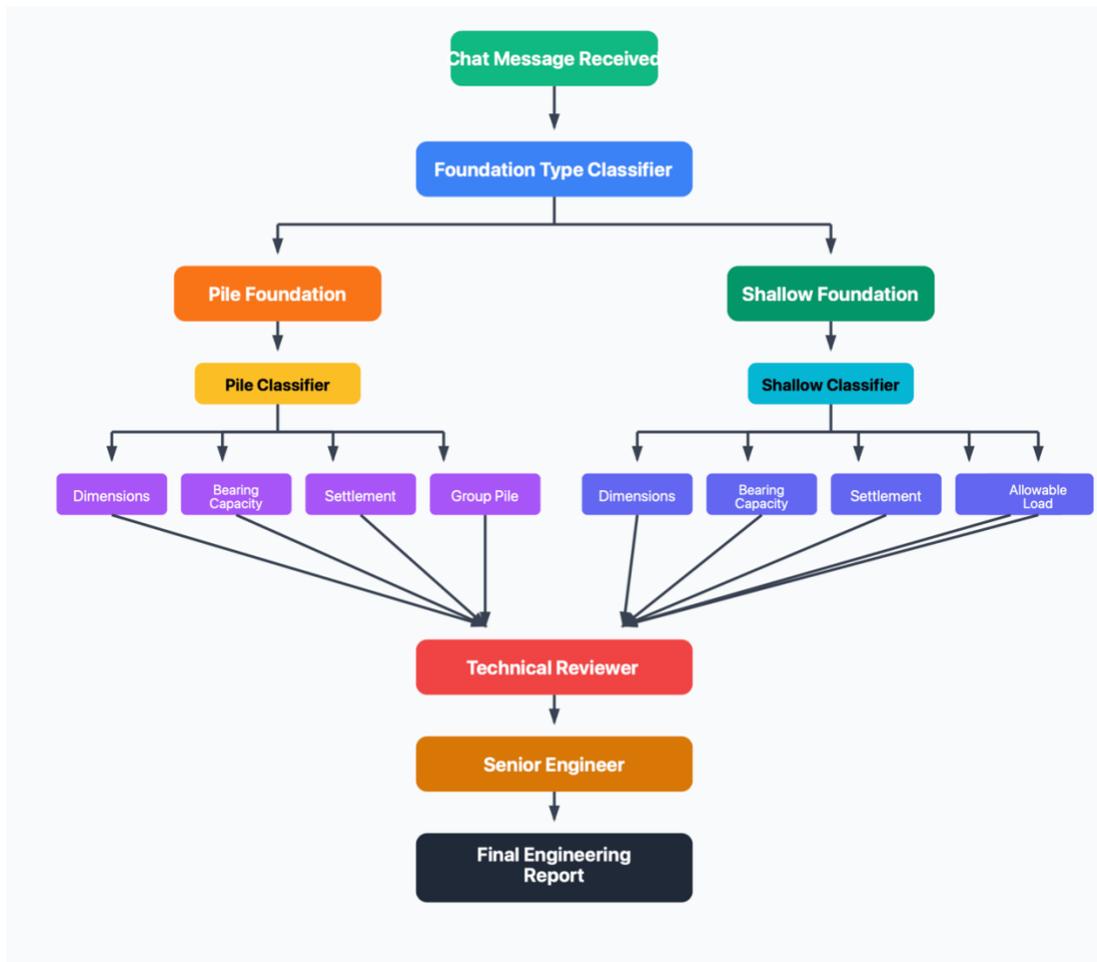

Fig.1 Agentic workflow of foundation design

This workflow diagram shown in Fig. 1 represents a comprehensive AI-driven geotechnical engineering system that addresses the growing need for automated foundation design tools in modern engineering practice. The system leverages natural language processing and machine learning classification to interpret engineering queries and route them through specialized computational modules, addressing the industry's challenge of standardizing complex geotechnical calculations while maintaining accuracy and code compliance. The initial classification stage employs text analysis algorithms to distinguish between pile and shallow foundation problems, which is critical given that these foundation types require fundamentally different analytical approaches. For pile foundations, the system addresses the four primary design considerations: dimensional optimization using methods like Vesic's method (Vesic, 1977) bearing capacity theories, ultimate bearing capacity calculations incorporating both skin friction and end-bearing components, settlement analysis through elastic and consolidation theories, and group pile effects that account for efficiency factors and block failure mechanisms. Similarly, shallow foundation analysis modules implement Terzaghi's bearing capacity theory and its modern extensions, addressing critical design parameters including foundation sizing based on allowable bearing pressure, ultimate bearing capacity calculations with appropriate shape and depth factors, immediate and consolidation settlement predictions, and safety factor verification against bearing capacity failure.

The multi-stage quality assurance framework embedded in this workflow addresses a critical gap in automated engineering systems by implementing human expert validation at two levels, ensuring compliance with professional engineering standards and liability requirements. The technical reviewer stage performs detailed verification of calculations against established geotechnical design codes, or local building codes, checking for computational accuracy, appropriate factor of safety applications, and adherence to soil mechanics principles. The senior engineer review provides final validation and professional formatting, ensuring that outputs meet the documentation standards required for engineering deliverables and regulatory approval processes. The integration of web search capabilities allows real-time access to updated design codes, standards, and technical references, addressing the dynamic nature of engineering regulations and best practices. This systematic approach to automated geotechnical design represents a significant advancement in engineering workflow optimization, potentially reducing design time while maintaining the rigor and safety standards essential for foundation engineering practice. The system's architecture supports scalability and customization for different regional codes and project types, making it adaptable to diverse engineering environments and regulatory frameworks while preserving the fundamental principles of safe and economical foundation design.

### 3.1 Agent Mathematical Formulation

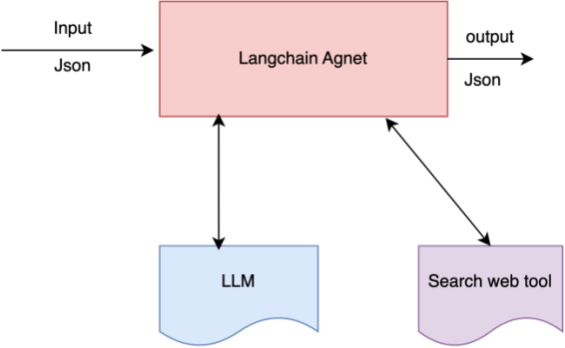

Fig 2. Component of AI agent workflow

Fig. 2 illustrates the mathematical formulation of the multi-agent system architecture, wherein each AI agent comprises a Large Language Model (LLM) functioning as the cognitive processing unit. The system accepts JSON-formatted input data and generates corresponding JSON output through the LangChain Agent framework. The architecture incorporates two primary computational components: the LLM module for core reasoning operations and a web search tool for external knowledge retrieval. In this implementation, SerpAPI (SerpApi, 2024) serves as the web search interface, enabling agents to access real-time information when the LLM's training data proves insufficient for specific design requirements. This hybrid approach addresses the inherent knowledge limitations of pre-trained language models while maintaining computational efficiency. The bidirectional information flow between the LangChain Agent and its constituent modules facilitates dynamic knowledge augmentation, wherein the system can autonomously determine when external data retrieval is necessary to complete design tasks. This architectural configuration enables the integration of static model knowledge with dynamic web-based information sources, thereby enhancing the system's capability to handle contemporary engineering problems that may not be adequately represented in the LLM's training corpus. Each AI agent $A_i$ in the system can be formally defined as:

$$A_i = \{LLM_{\{\theta\}}, P_i, T_i, K_i\} \quad (1)$$

where $LLM_{\{\theta\}}$ represents the parameterized language model with weights $\theta$, $P_i$ denotes the domain-specific prompt template for agent $I$, $T_i$ represents the tool set available to agent $I$, $K_i$ encompasses the knowledge base and constraints for domain $i$, The agent decision function is formulated as:

$$f_{i(x)} = LLM_{\{\theta\}(P_i \oplus x \oplus K_i)} \quad (2)$$

where $x$ represents the input query, $\oplus$ denotes prompt concatenation, and $f_{i(x)}$ produces the agent's response.

### 3.1 LangChain Tool Integration

The system integrates multiple tool types through LangChain's tool interface:

1. **Text Classifiers** ($C_{\{text\}}$): Implement hierarchical classification using embedding similarity
2. **Search Tools** ($T_{\{search\}}$): SerpAPI integration for real-time engineering standards retrieval

The tool invocation mechanism follows:

$$T_{output} = \sum_{j=1}^{n} w_j \cdot T_{j(context, parameters)} \quad (3)$$

where $w_j$ represents the confidence weight for tool $j$.

### 3.2 Classification Framework

The workflow initiates with a classification node that categorizes the design proposal and routes the task or prompt to the pre-configured agent designated for that specific task category. The dual-tier classification system employs transformer-based text classification with the following probability distribution:

$$P(class|query) = softmax(W \cdot LLM_{embed}(query) + b) \quad (4)$$

where $P(class|query)$ is the probability distribution over classes given a query, $LLM_{embed}(query)$ generates an embedding vector for the input query using a large language model, $W$ is the weight matrix that transforms the embedding to class logits and $b$ is the bias vector.

For pile foundations, the classification encompasses:

- **Pile Length**: Optimization of geometric parameters
- **Pile Bearing Capacity**: Load-bearing capacity assessment
- **Pile Settlement**: Displacement analysis under loading
- **Group Pile**: Multi-pile system analysis

For shallow foundations, the classification includes:

- **Foundation Dimensions**: Sizing requirements
- **Bearing Capacity**: Ultimate and allowable capacity determination
- **Settlement Analysis**: Immediate and consolidation settlement
- **Factor of Safety**: Risk assessment and reliability analysis
- **Allowable Load**: Service load determination

### 3.3 LLM-Enhanced Pile Foundation Analysis Agents

The calculation methodology is embedded as the system prompt for each agent node, providing the advantage that engineers can specify their preferred design approach or implement local design codes within the computational framework. Each agent incorporates not only calculation instructions but also employs one-shot learning techniques that provide the LLM with representative examples to enhance computational accuracy and consistency. This flexibility enables the system to accommodate varying international standards and regional engineering practices while maintaining computational rigor and enabling precise quantification of each settlement mechanism's contribution to overall pile performance.

The pile bearing capacity agent utilizes advanced prompt engineering that integrates geotechnical domain knowledge with mathematical computation capabilities. The agent architecture incorporates contextual information, mathematical formulations, worked examples, and validation criteria within a structured template framework. The system employs adaptive analytical method selection, utilizing Vesic's bearing capacity theory (Vesic, 1977) for deep foundation analysis. The settlement analysis agent employs a multi-component approach that calculates settlement contributions separately for shaft friction, end-bearing at the pile tip, and elastic compression of the pile itself. This methodology enables comprehensive settlement evaluation through the superposition of individual settlement components, where total pile settlement comprises the cumulative effects of shaft friction settlement along the pile length, settlement from end-bearing resistance at the pile tip, and elastic compression of the pile material itself. The shaft friction settlement component considers the load transfer mechanism along the pile-soil interface, while the tip settlement analysis evaluates the bearing response of soil beneath the pile base. The elastic compression of the pile is computed based on the applied load distribution and pile material properties. The group pile analysis agent implements a collaborative reasoning framework employing multiple LLM instances to evaluate different failure mechanisms. A consensus mechanism weights multiple evaluation perspectives to determine optimal group pile configurations, incorporating group efficiency calculations through established geotechnical methodologies. The example of system prompt is shown in Fig. 3.

```
1. Calculating Ultimate Bearing Capacity (Single Pile)

Objective: Compute the ultimate bearing capacity (( Q_u )) of a single pile, which is the sum of end-bearing (( Q_p )) and skin friction (( Q_s )) capacities, using Vesic's method.

Steps:
Define the Problem:
Identify the soil profile, including layer thicknesses and properties (e.g., ( s_u ) for clay, ( \phi ) for sand, unit weight ( \gamma )).

Specify pile type (e.g., driven concrete), diameter (( D )), and length (( L )).

Note groundwater table depth and any additional data like ( E_s ).

-End-bearing:
  -Clay: ( Q_p = A_p \cdot s_u \cdot N_c ), where ( N_c = 9.0 ).
  -Sand: ( Q_p = A_p \cdot \sigma_v' \cdot N_q ), where ( N_q = e^{\pi \tan \phi} \tan^2 \left(45^\circ + \frac{\phi}{2}\right) ).
( A_p = \frac{\pi D^2}{4} ).

Skin friction:
-Clay: ( f_s = \alpha s_u ).
-Sand: ( f_s = K \sigma_v' \tan \delta ).
( Q_s = \sum (f_s \cdot A_s) ), where ( A_s = \pi D h_i ).
Vesic's method (if ( E_s ) available):
Adjust ( N_q ):

Calculate Effective Stress (( \sigma_v' )):
For each layer:
Above groundwater: ( \sigma_v' = \gamma h_i ).
Below groundwater: ( \sigma_v' = (\gamma - \gamma_w) h_i ).
Sum stresses to the pile tip for ( Q_p ).
For ( Q_s ), use average ( \sigma_v' ) in each layer: ( \sigma_v' = \frac{\sigma_{v,\text{top}}' + \sigma_{v,\text{bottom}}'}{2} ).
Compute End-Bearing Capacity (( Q_p )):
Determine the soil at the pile tip.
Apply the appropriate formula based on soil type.

Compute Skin Friction Capacity (( Q_s )):
For each layer, calculate ( f_s ) and ( A_s ).
Sum contributions: ( Q_s = \sum Q_{s,i} ).
Calculate Ultimate Bearing Capacity:

( Q_u = Q_p + Q_s ).
Present Results:
Provide ( Q_u ) in kN.
If requested, compute ( Q_a = \frac{Q_u}{FS} ).
```

**Fig. 3** Example of system prompt

### 3.4 LLM-Enhanced Shallow Foundation Analysis Agents

The adaptive bearing capacity agent incorporates dynamic prompt adaptation capabilities that respond to varying soil conditions and foundation geometries through a multi-tiered computational framework as shown in Fig. 4. The system employs a hierarchical agent architecture where task classification directs geotechnical problems to specialized modules, followed by a calculation agent implementing role-based prompt engineering with embedded geotechnical engineer persona and systematic methodological instructions based on Terzaghi's bearing capacity theory. The prompt structure integrates one-shot learning examples to guide large language model (LLM) reasoning in step-wise calculation procedures, enabling comprehensive bearing capacity and settlement analysis including immediate settlement and consolidation settlement components for normally consolidated clay conditions. An exemple question prompt structure is illustrated in Fig. 5, demonstrating the systematic approach to problem formulation and solution methodology. This approach demonstrates effective integration of machine learning capabilities with traditional geotechnical engineering principles, where LLM-informed prior distributions enhance uncertainty quantification reliability in settlement predictions while maintaining adherence to established engineering methodologies and providing enhanced analytical capabilities for foundation design applications.

You are a professional geotechnical engineer trained in soil mechanics, foundation design, and site investigations. Your objective is to provide accurate, structured, and justifiable engineering calculations in compliance with appropriate geotechnical terminology and standard. Design foundations following these steps:
1. Start with a problem summary. Clearly define what is being calculated (e.g., design bearing capacity of a shallow foundation).
2. Identify key inputs and assumptions (e.g., soil properties, geometry, design approach used).
3. Break down the problem into clear, step-by-step calculations.
4. When calculating for factors, make sure to use formulas and not the tables or charts.
5. Show how each formula is applied using the given values
6. Provide numerical results with proper units
7. Summarize the final answer at the end

Avoid using phrases like "it depends" or "you should consult a table." Always provide clear chain of thoughts with engineering reasoning for each step and avoid shortcuts or assumptions not explicitly stated in the input. If the problem lacks enough information, explain what additional data would be required for a complete calculation.
If you are not sure please search web site.

----

You are the agent for the calculation of Shallow Foundation Dimensions Following Terzaghi's Bearing Capacity Theory.

**Objective:**
To determine the required dimensions (width and length) of a shallow foundation using Terzaghi's bearing capacity theory. This approach can be applied flexibly to square, rectangular, and strip foundations on cohesive or cohesionless soils.

---

**1. General Input Parameters for Various Question Types:**

* Depth of foundation ($D_f$)
* Water table depth (for submerged conditions)
* Unit weight ($\gamma$) and saturated unit weight ($\gamma_{sat}$)
* Effective friction angle ($\emptyset'$)
* Effective cohesion ($c'$)
* Applied vertical load (Q or V)
* Applied moment (if any) (M)
* Allowable bearing pressure ($q_{allow}$)
* Factor of safety (FS)
* Width-to-length ratio (B/L) if given
* Shape of foundation: square, rectangular, or strip

---

**2. Terzaghi's General Bearing Capacity Equation:**

$q_{ult} = c*N_c*s_c*d_c + q*N_q*s_q*d_q + 0.5*\gamma*B*N_\gamma*s_\gamma*d_\gamma$

Where:

* $q_{ult}$ = ultimate bearing capacity (kN/m$^2$)
* c = cohesion (kPa)
* q = effective overburden pressure at foundation base (kPa)
* B = foundation width (m)
* $N_c, N_q, N_\gamma$ = bearing capacity factors (based on $\emptyset'$)
* $s_c, s_q, s_\gamma$ = shape factors (based on B/L ratio)
* $d_c, d_q, d_\gamma$ = depth factors (optional; use 1.0 unless advanced analysis)

---

**3. Determine Effective Overburden Pressure (q):**

If groundwater is above the foundation base:
q = γ * z + γ' * (Df - z)
Where:

* z = depth to water table
* γ' = $\gamma_{sat}$ - $\gamma_w$ ($\gamma_w$ = 9.81 kN/m$^3$)

If water table is below the base:
q = γ * Df

---

**4. Determine Foundation Area Based on Allowable Bearing Pressure:**

If $q_{allow}$ is given:
A = (V ± M/e) / $q_{allow}$
Where:

* V = vertical load (DL + LL)
* M = applied moment (kN•m)
* e = eccentricity = M / V
* A = required area = B × L
* e < B/6 and e < L/6 to avoid tension (use effective area if e > B/6)

---

Fig. 4 Example system prompt for shallow foundation design

```
Example of question:
#Find the footing dimensions B x L to carry a moment induced by winds of 800 kN.m.
The DL is 800 kN, the LL is 800 kN, and the allowable soil pressure qallow = 200 kPa, and FS = 2.5.

#Find the footing dimensions B x L to carry a moment induced by winds of 1500
kN.m. The DL is 1200 kN, the LL is 1000 kN, and the allowable soil pressure qallow
= 200 kPa, and FS = 2.5.

#A concrete pile is 15 m long with 0.45 m × 0.45 m cross-section
Embedded in sand having γ = 17 kN/m³ and φ = 35°
Calculate:
1. Ultimate skin friction (Q_s) (K=1.3, δ'=0.8φ')
2. Ultimate bearing capacity (Q_all) with safety factors 3

#Find the penetration depth for a 35cm square R.C pile driven through seabed to
carry: Max compressive load: 500 kN, Net uplift load: 300 kN
Soil layers:1. 10m saturated medium dense sand (SPT=12). Dense sand-gravel
(SPT=40). Safety factors: 2.5 (uplift), 4 (compression)
```

Fig. 5 The Example of question

### 3.5 Technical Reviewer Agent

The reviewer agent implements a multi-criteria evaluation framework utilizing a weighted scoring mechanism to assess generated content quality. The system prompt of reviewer is shown in Fig. 6. The scoring function is formally defined as:

$$Score_{quality} = \sum_{i=1}^{m} w_i \cdot C_i(calculation, standards) \tag{5}$$

where $C_i$ represents individual quality criteria and $w_i$ denotes importance weights learned through reinforcement learning from expert feedback.

```
You are geotechnical engineer expert check the calculation and revise it.

1. Carefully evaluate the design against geotechnical standards and code requirements

2. Provide a detailed review with specific technical feedback

3. recheck calculation

4. Include a quality assessment score from 1-10

5. Explicitly state whether the design PASSES or REQUIRES REVISION
```

**Fig. 6** The system prompt of reviewer

Upon computation of the aggregate score, the reviewer agent performs a dual function: (1) generating revised content based on identified deficiencies across the evaluated criteria, and (2) propagating both the calculated score and revised output to the downstream senior engineering node. This bifurcated workflow ensures that subsequent processing stages receive both quantitative assessment metrics and qualitatively improved content, facilitating informed decision-making in the final conclusion phase. The weighted aggregation approach maintains interpretability of the scoring mechanism while enabling systematic quality enhancement through the multi-agent pipeline

### 3.6 Senior Engineer Agent

The senior engineer agent represents the terminal node in the multi-agent pipeline, receiving comprehensive feedback comprising the reviewer's quantitative assessment scores and qualitatively enhanced content. The system prompt is shown in Fig. 7. This agent implements a hierarchical decision tree framework encoded directly within the large language model architecture, enabling sophisticated reasoning over multiple input modalities.

---
You are senior geotechnical engineer. Review the calculation and create design report. Check the design step to be design code of practice.

-Formatting report in simple text format and clearly present the final answer.

---

Fig. 7 The system prompt of senior engineer

### 4. Experiment

Table 2 presents a structured evaluation framework for assessing AI model performance in geotechnical engineering applications using a four-point scale ranging from 0.0 to 2.0 points across four critical evaluation dimensions. While computational accuracy represents a fundamental requirement, relying solely on accuracy metrics provides a dangerously incomplete assessment of AI model performance in safety-critical geotechnical applications (Baghbani et al., 2022; Guo et al., 2023). The evaluation framework deliberately incorporates multiple dimensions because accurate final results can emerge from fundamentally flawed reasoning processes, creating a false sense of reliability that could lead to catastrophic failures in real-world applications. This risk is particularly pronounced given that soils and rocks exhibit complex behaviors and high uncertainty in material modeling, making traditional physically-based engineering approaches insufficient. A multi-dimensional hierarchical evaluation system is essential to estimate data quality and AI model performance, as multiple key dimensions are needed to evaluate specific conditions separately, allowing for a clearer understanding of strengths and weaknesses among various evaluation criteria (Zhang et al., 2024).

The chain-of-thought reasoning criterion (Wei et al., 2022) becomes absolutely critical as it exposes the internal logic pathways that lead to solutions, revealing whether the model truly understands underlying engineering principles or merely produces correct answers through computational coincidence. Without examining the reasoning process, engineers cannot identify when models apply inappropriate methodologies that happen to yield correct results for specific cases but would fail under different conditions. The complex scenario handling dimension measures model robustness when confronted with non-standard boundary conditions and multifaceted geotechnical problems, exposing vulnerabilities that might remain hidden when models are evaluated only on straightforward calculations (Khedher et al., 2023). The increasing use of artificial neural networks in safety-critical systems requires ensuring robustness against out-of-distribution shifts in operation, where consequences of systemic failures can result in loss of life, economic disruption, or environmental harm (Yousefpour et al., 2025). Professional standards organizations have developed comprehensive frameworks for AI system evaluation, with IEEE standards such as IEEE 3168-2024 (IEEE, 2024) for robustness evaluation establishing measurable and testable levels of transparency for autonomous systems.

The consistent structured output format criterion evaluates output standardization and organizational consistency, directly impacting practical usability and professional acceptability of AI-generated solutions in engineering practice. This standardization becomes particularly critical when AI systems must interface with existing engineering workflows, databases, and regulatory compliance systems, as structured evaluation can provide clear, interpretable results that explain how and why a model behaves in specific ways (Liang et al., 2025). This multi-dimensional approach recognizes that in geotechnical engineering, where human safety and structural integrity are paramount, engineers must understand not just what the AI calculates, but how it thinks, why it selects specific approaches, and whether its reasoning remains valid across diverse scenarios that extend beyond the training data (Baghbani et al., 2022). The potential for machine learning to shape geotechnical engineering practice is immense, but the agenda should not focus on applying algorithms alone—the geotechnical context that gives rise to the data is critical, with physics-informed results being explainable and interpretable. This comprehensive evaluation framework ensures that AI models deployed in geotechnical engineering applications meet the rigorous standards necessary for safety-critical infrastructure projects, addressing the unique challenges posed by the inherent uncertainties and complex behaviors characteristic of geotechnical materials and systems. The framework represents a foundational step toward establishing standardized, multi-dimensional assessment protocols that can support the responsible integration of AI technologies into professional geotechnical engineering practice, ultimately contributing to safer, more reliable, and more efficient infrastructure development worldwide.

**Table 2** Evaluation Rubric for Model Performance Assessment in Geotechnical Engineering Applications

| # | Criteria | Excellent (2.0 points) | Good (1.0 points) | Needs Improvement (0.5 points) | Poor (0.0 points) |
|---|---|---|---|---|---|
| 1 | Accuracy Calculations | The model provides completely accurate calculations for all requirements, including bearing capacity, slope stability, and foundation design. | The model provides mostly accurate calculations but may have minor errors in complex scenarios. | The model struggles with accuracy, particularly in complex scenarios. | The model fails to provide accurate calculations. |
| 2 | Chain-of-Thought (CoT) Reasoning | The model demonstrates fully correct and logical reasoning steps, clearly explaining how it arrives at each calculation or design decision. | The model provides mostly correct reasoning but may have minor logical flaws or omissions. | The model shows some reasoning ability, but the logic is flawed or incomplete. | The model fails to provide coherent reasoning or skips critical steps in the calculation process. |
| 3 | Handling Complex Scenarios | The model excels in handling complex geotechnical scenarios, with simplified solutions, clear explanations, and non-standard boundary conditions. | The model performs well in most scenarios but may struggle with highly complex or unusual cases. | The model can handle simple scenarios but struggles in moderately complex cases. | The model fails to handle even basic geotechnical scenarios. |
| 4 | Consistent & Structured Output Format | The model consistently follows a structured output format, clearly presenting inputs, calculations, and final results. | The model mostly follows a structured format but may have occasional inconsistencies. | The model requires manual adjustments to maintain a structured output. | The model produces unstructured or inconsistent output. |

The systematic removal of the router multi-agent component serves multiple analytical objectives within the experimental framework. Primary objectives include quantifying the router's contribution to task completion efficiency, solution quality metrics, and computational resource optimization. Secondary analysis focuses on identifying potential performance bottlenecks, failure modes, or inefficiencies introduced by complex routing mechanisms compared to simplified sequential processing approaches. The comparative methodology enables empirical validation of the multi-agent architecture's effectiveness while providing quantitative insights into the optimal balance between system complexity and measurable performance gains across diverse task categories and operational scenarios. To achieve these analytical objectives, the experimental design implements three distinct agentic workflow configurations, each representing a systematic reduction in architectural complexity. The baseline configuration maintains the complete multi-agent ecosystem as illustrated in the provided architectural diagram, featuring the full integration of specialized agents with dynamic routing capabilities. This comprehensive system orchestrates task flow through the Designer agent for initial problem decomposition and solution conceptualization, the Reviewer agent for iterative quality assessment and validation procedures, and the Senior Engineer agent for final implementation synthesis and optimization protocols. Each agent operates with dedicated computational resources, accessing distinct large language model instances through OpenRouter (OpenRouter, 2024) Chat Model configurations and specialized API integrations including SerpAPI for enhanced information retrieval capabilities.

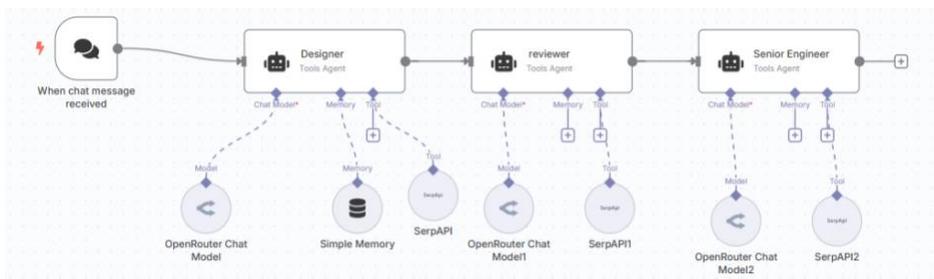

Fig. 8 The agentic workflow type I

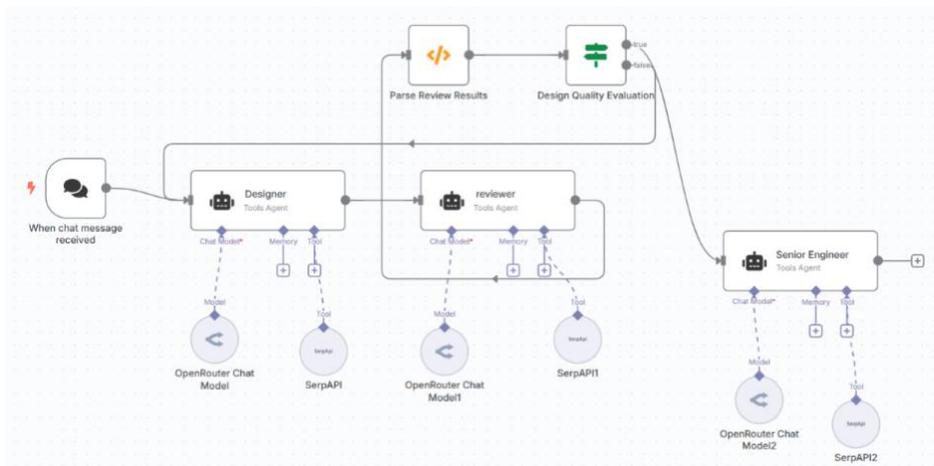

Fig. 9 The agentic workflow type II

The first ablated configuration (Fig. 8) strategically eliminates the router multi-agent subsystem while preserving core agent functionality, transforming the dynamic routing system into a static sequential pipeline. This modification removes adaptive task distribution, intelligent load balancing, and context-aware routing decisions, enabling precise measurement of routing intelligence contributions to system performance. The second agentic workflow (Fig. 9) introduces an iterative refinement mechanism centered on the Reviewer agent's quality control capabilities. The Reviewer agent functions as a quality gatekeeper with authority to reject, validate, or request revisions from both Designer and Senior Engineer agents. The workflow implements feedback loops where the Reviewer agent evaluates deliverables against predefined quality criteria and redirects tasks for iterative refinement when deficiencies are identified.

The evaluation protocol encompasses twenty-seven distinct test cases spanning seven primary categories of geotechnical engineering problems. Shallow foundation analysis includes fifteen test cases covering dimensional design calculations, allowable load determinations, bearing capacity computations, and settlement analysis. Pile foundation evaluation extends this framework with twelve additional test cases across four specialized categories: dimensional design calculations, bearing capacity analysis including skin friction and ultimate bearing calculations, settlement computations for prestressed concrete piles, and pile group analysis encompassing load distribution and elastic settlement behavior. Each test case was executed three times to account for the stochastic nature of large language model outputs and quantify variability in model responses. This triple-trial approach enables comprehensive analysis of model stability, output consistency, and performance reliability under identical input conditions. The experimental design maintains controlled testing environments where each test case preserves identical input conditions across all trial iterations, isolating the impact of model randomness on output quality and accuracy. The evaluation framework systematically captures multiple performance indicators including numerical accuracy, output format consistency, computational methodology validity, and frequency of invalid outputs. This systematic ablation methodology enables comprehensive evaluation of each architectural component's contribution to overall system performance through comparative analysis across task completion accuracy, solution quality metrics, and computational reliability characteristics.

## 5. Results of Model Evaluation

The investigation examines the efficacy of routing-based agentic selection mechanisms through comparative analysis of multiple AI model configurations across specialized engineering assessment criteria. Dataset I (Table 3), focused on shallow foundation design, encompassed five critical parameters: Finding Dimensions, Allowable Load, Bearing Capacity, Settlement, and Factor of Safety, while Dataset II Table (4), dedicated to pile design applications, evaluated four specialized criteria: Finding Dimensions, Bearing Capacity, Settlement, and Pile Group analysis. Nine model configurations were systematically evaluated, including baseline large language models (Deepseek R1, Gemini 2.5 Pro-preview, ChatGPT 4.0-turbo, Grok 3), conventional agentic workflows (Workflows I and II), and the proposed routing-based agentic selection framework. The routing-based system implements dynamic model selection mechanisms that optimize task allocation based on problem characteristics and historical performance patterns, incorporating three key architectural components: dynamic task allocation algorithms, heterogeneous model integration protocols, and adaptive learning frameworks for continuous optimization of routing strategies.

Table 3 The performance results of different type of large language model for design of shallow foundation

| MODELS | Finding Dimensions | Allowable Load | Bearing Capacity | Settlement | Factor of Safety | Average Grade (%) |
|---|---|---|---|---|---|---|
| | Grade (%) | Grade (%) | Grade (%) | Grade (%) | Grade (%) | |
| Agentic Workflow I with Gemini 2.5 pro | 50.00 | 31.25 | 43.75 | 31.25 | 37.50 | 38.75 |
| Agentic Workflow II with Gemini 2.5 pro | 62.50 | 50.00 | 50.00 | 100.00 | 100.00 | 72.50 |
| **Proposed agentic workflow with Gemini 2.5 pro** | 87.50 | 87.50 | 75.00 | 87.50 | 75.00 | 82.50 |
| Deepseek R1 | 75.00 | 87.50 | 75.00 | 75.00 | 75.00 | 77.50 |
| Gemini 2.5 Pro-preview | 87.50 | 62.50 | 75.00 | 62.50 | 68.75 | 71.25 |
| ChatGPT 4.0-turbo | 62.50 | 56.25 | 75.00 | 43.75 | 56.25 | 58.75 |
| Grok 3 | 100.00 | 87.50 | 100.00 | 75.00 | 68.75 | 86.25 |
| Agentic Workflow I - with GROK 3 | 87.5 | 56.25 | 87.5 | 62.5 | 62.5 | 71.25 |
| Agentic Workflow II - with GROK 3 | 75 | 56.25 | 62.5 | 87.5 | 87.5 | 73.75 |
| **Proposed agentic workflow - with GROK 3** | 100 | 87.5 | 100 | 87.5 | 100 | 95.00 |

Bold is the best performance

Table 4 The performance results of different type of large language model for design of pile foundation

| MODELS | Finding Dimensions | Bearing Capacity | Settlement | Pile Group | Average Grade (%) |
|---|---|---|---|---|---|
| | Grade (%) | Grade (%) | Grade (%) | Grade (%) | |
| Agentic Workflow I with Gemini 2.5 pro | 43.75 | 56.25 | 37.50 | 68.75 | 51.56 |
| Agentic Workflow II with Gemini 2.5 pro | 50.00 | 37.50 | 75.00 | 62.50 | 56.25 |
| Proposed agentic workflow with Gemini 2.5 pro | 87.50 | 75.00 | 75.00 | 100.00 | 84.38 |
| Deepseek R1 | 87.50 | 62.50 | 62.50 | 87.50 | 75.00 |
| Gemini 2.5 Pro-preview | 75.00 | 87.50 | 62.50 | 100.00 | 81.25 |
| ChatGPT 4.0-turbo | 87.50 | 62.50 | 62.50 | 87.50 | 75.00 |
| Grok 3 | 100.00 | 87.50 | 75.00 | 87.50 | 87.50 |
| Agentic Workflow I - with GROK 3 | 43.75 | 62.5 | 50 | 50 | 51.56 |
| Agentic Workflow II - with GROK 3 | 75 | 50 | 62.5 | 62.5 | 62.5 |
| **Proposed agentic workflow with Grok 3** | **87.50** | **87.50** | **87.50** | **100.00** | **90.63** |

Bold is the best performance

    The routing-based agentic workflow achieved superior performance across both shallow foundation and pile design applications, demonstrating the versatility and effectiveness of intelligent model orchestration in specialized geotechnical domains. With Grok 3 as the foundational model, average performance scores of 95.00% for shallow foundation design and 90.63% for pile design were obtained, representing improvements of 8.75 and 3.13 percentage points over standalone Grok 3 performance, respectively. When implemented with Gemini 2.5 Pro, the routing-based approach yielded 82.50% average performance for shallow foundation design and 84.38% for pile design applications, substantially exceeding conventional workflow performance by margins ranging from 10.0 to 43.75 percentage points. Cross-domain analysis revealed consistent performance advantages for the routing-based framework across all engineering assessment criteria, with particular excellence in specialized calculations where domain expertise is critical. In shallow foundation Finding Dimensions calculations, the proposed system achieved perfect performance (100%) with Grok 3 and maintained 87.50% performance with Gemini 2.5 Pro, compared to variable conventional workflow performance ranging from 43.75% to 87.50%. Similarly, in pile design applications, the routing system demonstrated robust performance with 87.50% in Finding Dimensions calculations using Grok 3, while conventional workflows exhibited substantial performance degradation across both foundational models.

    The superior performance of the routing-based agentic workflow in both shallow foundation and pile design applications can be attributed to its ability to dynamically assess problem complexity specific to each geotechnical domain and implement optimal model selection strategies tailored to the computational requirements of foundation engineering versus deep foundation systems. The consistency of performance improvements across these distinct engineering specializations, with only 4.37% and 1.88% variation between datasets for Grok 3 and Gemini 2.5 Pro implementations respectively, demonstrates the robustness of the routing mechanism in adapting to different geotechnical computational requirements while maintaining optimization strategies. In highly specialized domains such as Pile Group analysis, which requires complex understanding of group effects, load distribution, and soil-structure interaction, the routing approach achieved perfect performance (100%) with both foundational models, while conventional workflows demonstrated significant performance degradation, with scores ranging from 50.00% to 68.75%, indicating the critical importance of intelligent task allocation in complex geotechnical calculations.

    Conventional Agentic Workflow I consistently underperformed across all experimental conditions in both shallow foundation and pile design applications, achieving average performance scores of 38.75% and 51.56% with Gemini 2.5 Pro across the respective datasets, with similarly poor performance using Grok 3 (71.25% and 51.56%). This systematic underperformance across distinct geotechnical domains indicates fundamental architectural limitations in sequential processing approaches that fail to leverage the specialized computational strengths required for different foundation design methodologies. Agentic Workflow II demonstrated improved but

inconsistent performance relative to Workflow I, with average scores varying significantly between shallow foundation design (72.50% with Gemini 2.5 Pro) and pile design applications (56.25% with Gemini 2.5 Pro), suggesting that while this architecture achieves better task coordination than Workflow I, it lacks the adaptive optimization capabilities necessary for consistent cross-domain performance in specialized geotechnical engineering applications.

Individual model analysis revealed distinct computational strengths specific to shallow foundation versus pile design applications that the routing system successfully identified and leveraged through intelligent orchestration. Grok 3 demonstrated exceptional performance in geometric and structural analysis tasks across both domains, achieving perfect performance in Finding Dimensions for shallow foundation design and maintaining high performance (87.50%) in pile design applications, with particular strengths in Bearing Capacity calculations where it achieved 100% and 87.50% performance respectively. Gemini 2.5 Pro-preview exhibited domain-specific advantages that varied between foundation types, achieving 75.00% performance in shallow foundation Bearing Capacity analysis compared to 87.50% in pile design Bearing Capacity calculations, and demonstrating perfect performance in specialized Pile Group analysis (100%), indicating the value of heterogeneous model integration in routing-based architectures for capturing the nuanced computational requirements of different foundation design methodologies.

The experimental results establish routing-based agentic selection as a fundamental advancement in AI orchestration methodologies for geotechnical engineering applications, with the 8.75 to 43.75 percentage point performance improvements observed across shallow foundation and pile design domains representing statistically significant enhancements that justify the increased architectural complexity of routing-based systems. The consistent performance advantages across different foundational models and both geotechnical specializations indicate exceptional robustness and demonstrate the framework's ability to adapt to the distinct computational challenges inherent in shallow versus deep foundation design, including the complex soil-structure interaction analyses required for pile group behavior versus the bearing capacity and settlement calculations critical for shallow foundation systems. These findings suggest that future agentic AI development should prioritize dynamic routing mechanisms over conventional sequential workflows, particularly for specialized geotechnical domains where the integration of multiple computational approaches and the intelligent selection of appropriate analytical methods are essential for accurate engineering analysis. The routing-based framework's superior performance in both shallow foundation and pile design applications provides a scalable approach for addressing the diverse computational challenges across the broader spectrum of geotechnical engineering practice, establishing it as a promising direction for future agentic AI development in foundation engineering and related technical domains.

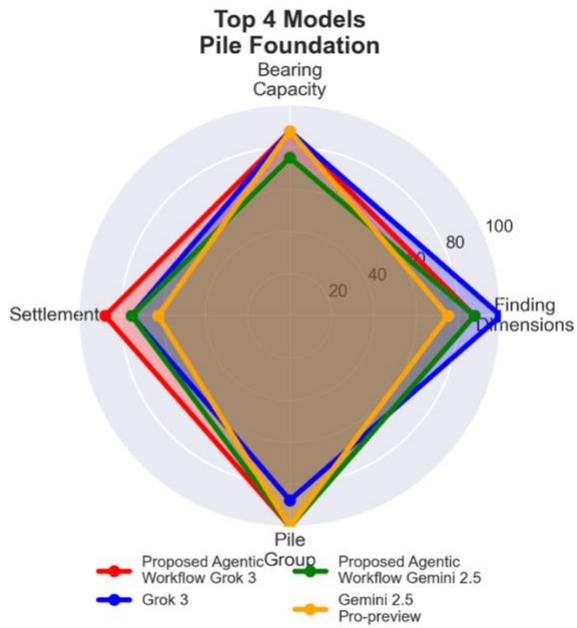

Figure 10 Comparison top 4 model in pile foundation design

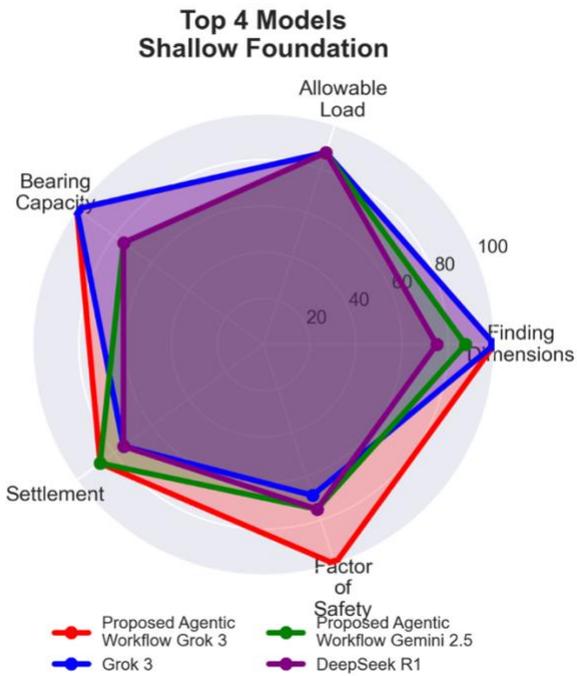

Figure 11 Comparison top 4 model in shallow foundation design

The radar charts in Figures 10 and 11 provide visual analysis of the four highest-performing

AI model configurations, validating the superiority of the proposed routing-based agentic workflow architecture. In pile foundation design (Figure 10), the proposed agentic workflow with Grok 3 demonstrates exceptional performance consistency, achieving near-perfect scores across all four assessment criteria. Grok 3 standalone exhibits strong performance in Finding Dimensions and Bearing Capacity but shows reduced effectiveness in Settlement and Pile Group analyses, limitations effectively compensated by the routing framework.The shallow foundation design analysis (Figure 11) reveals similar patterns, with the proposed agentic workflow with Grok 3 maintaining superior performance across five assessment criteria, particularly excelling in Allowable Load and Bearing Capacity calculations. Grok 3 standalone shows exceptional performance in Finding Dimensions but reduced effectiveness in Factor of Safety calculations, successfully addressed by the routing framework. Cross-domain comparison reveals pile foundation design shows more pronounced performance variations between models, particularly in Settlement and Pile Group analyses, reflecting the increased complexity of deep foundation systems requiring sophisticated understanding of group effects and soil-pile interaction behaviors. The geometric patterns validate the routing mechanism's effectiveness, creating enlarged performance envelopes that encompass and exceed individual model capabilities, particularly evident in complex criteria where standalone models demonstrate limitations. These visualizations demonstrate that the routing-based agentic workflow represents a fundamental advancement in AI-assisted foundation design, providing superior performance through intelligent orchestration while adapting to specific requirements of different foundation engineering specializations.

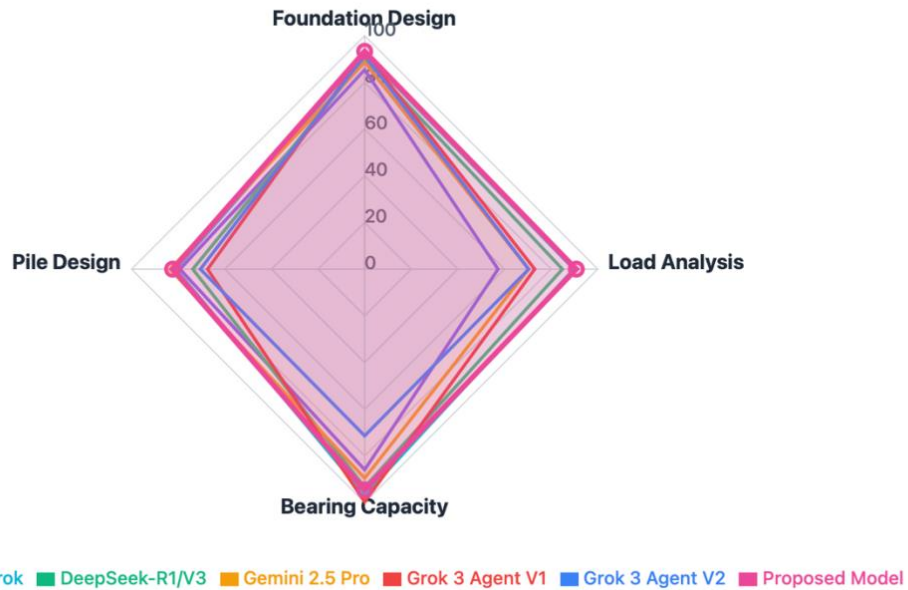

Fig. 12 The radar analysis of different model for accuracy

The radar chart in Fig. 12 provides quantitative performance assessment of seven AI models across four primary foundation engineering task categories, utilizing accuracy derived from mean absolute percentage error (MAPE) calculations with radial distance from center (0% accuracy) to periphery (100% accuracy) representing computational precision. Performance analysis reveals heterogeneous capabilities across Foundation Design where ChatGPT 4.0 exhibits minimal accuracy, base Grok demonstrates improved performance, and the Proposed Model Agent w/ Grok 3 maintains comparable accuracy to base Grok, while Load Analysis presents the most pronounced variability with multiple models including base Grok, DeepSeek-R1/V3, and Proposed Model Agent w/ Grok 3 achieving maximum accuracy compared to ChatGPT 4.0, Gemini 2.5 Pro, and Grok 3 Agent variants demonstrating complete computational failure (0% accuracy). Bearing Capacity calculations show consistent high-performance characteristics with base Grok achieving maximum accuracy and Proposed Model Agent w/ Grok 3 maintaining competitive performance, while the Pile Design axis exhibits the most significant performance differential where the Proposed Model Agent w/ Grok 3 demonstrates exceptional accuracy compared to all competing models, representing breakthrough computational capabilities in pile design applications. The Proposed Model Agent w/ Grok 3 exhibits an expanded polygonal configuration with maximum area coverage and optimal radial extension across multiple axes, particularly demonstrating exceptional performance in Pile Design while maintaining competitive accuracy in Load Analysis and Bearing Capacity domains, validating the proposed model's applicability for comprehensive foundation engineering workflows and representing an evolutionary advancement that preserves fundamental computational strengths while achieving superior performance in specialized geotechnical applications.

## 5. Discussion

The experimental results establish router-based multi-agent architectures as a significant advancement in AI-assisted geotechnical engineering applications, demonstrating superior performance across diverse foundation design tasks while revealing critical insights regarding current AI capabilities and future potential in safety-critical engineering domains.

### 5.1 Breakthrough in LLM Computational Capabilities

This study demonstrates a fundamental paradigm shift in large language model computational abilities, addressing a historically significant limitation that has constrained AI applications in quantitative engineering domains. Traditional LLMs have exhibited notorious weaknesses in mathematical calculations, frequently producing erroneous results even for elementary arithmetic operations, leading to widespread implementation of external calculation tools and computational APIs as auxiliary systems. Previous research has extensively documented these limitations: Chen et al. (2024) demonstrated that GPT-4 achieved only 28.9% accuracy with zero-shot learning in geotechnical engineering calculations, requiring extensive examples to reach 67% accuracy, while Xu et al. (2025) implemented RAG-based approaches with pre-built multimodal databases containing worked examples to compensate for models' inability to perform independent mathematical reasoning. Contemporary AI implementations have predominantly relied on hybrid architectures partitioning linguistic reasoning and mathematical computation into separate systems, utilizing tools such as LangChain calculator interfaces, Wolfram Alpha integration, and Python code execution environments specifically because the models themselves could not reliably perform quantitative analyses essential for engineering practice.

The exceptional performance observed in this study represents a revolutionary advancement in direct LLM computational capabilities. The router-based system with Grok 3 achieved 95.00% performance for shallow foundation design and 90.63% for pile design, while standalone Grok 3 demonstrated strong performance (86.25% for shallow foundations and 87.50% for pile foundations) across diverse geotechnical calculations, all without requiring external computational tools or extensive example-based prompting. This computational breakthrough is particularly significant given the complexity of geotechnical engineering calculations, which involve multi-step analytical procedures, integration of multiple design parameters, application of empirical factors, and compliance with established design codes. The strong performance achieved by Grok 3 across various calculation types—requiring understanding of soil mechanics principles, application of design factors, and integration of safety considerations—indicates that current LLMs have transcended basic arithmetic limitations to achieve genuine mathematical reasoning capabilities in specialized technical domains without dependence on external examples or computational assistance.

### 5.2 Performance Enhancement and Grok 3's Exceptional Capabilities

The router-based multi-agent system achieved exceptional performance scores of 95.00% for shallow foundation design and 90.63% for pile design, representing improvements of 8.75 and 3.13 percentage points over standalone Grok 3 performance (86.25% and 87.50%, respectively). This superior performance results from dynamic task allocation to specialized agents optimized for specific problem domains, leveraging distinct computational strengths while mitigating individual model weaknesses through complementary agent collaboration.The systematic outperformance of conventional agentic workflows by margins ranging from 10.0 to 43.75 percentage points validates the critical importance of intelligent task allocation. Notably, the router-based system achieved perfect performance (100%) in Pile Group analysis with both Grok 3 and Gemini 2.5 Pro implementations, representing a significant achievement given the complexity of pile group behavior analysis involving group efficiency factors and soil-structure interaction effects.

Grok 3's exceptional performance as a standalone model, demonstrating strong results across multiple geotechnical engineering domains without requiring extensive system prompt engineering, example-based instruction, or RAG retrieval systems, indicates emerging problem-solving abilities that approach characteristics expected of artificial general intelligence (AGI) in specialized technical domains. This autonomous computational capability suggests inherent understanding of mathematical reasoning and engineering principles that extends beyond pattern recognition, representing a crucial step toward artificial general intelligent (AGI) in engineering applications.

### 5.3 Cross-Domain Applications and Scalability

The router-based architecture demonstrates significant potential for extension across diverse engineering disciplines, particularly in structural and geotechnical design domains. The architectural principles—intelligent task classification, domain-specific expert agents, and adaptive routing mechanisms—provide a scalable foundation for addressing complex engineering calculations across multiple specialized domains without requiring extensive example databases or external computational tools.

For reinforced concrete design, the framework could distinguish between structural element types and loading conditions, enabling specialized agent deployment for flexural design, shear design, deflection analysis, and seismic design considerations. Each agent would incorporate specific design codes while maintaining specialized knowledge of material behavior and failure modes. Beyond foundation design, the architecture shows exceptional promise for comprehensive geotechnical engineering applications including retaining wall design, slope stability analysis, and ground improvement design.

The modular agent structure enables domain-specific specialization without compromising system-wide performance, allowing individual agents to maintain deep expertise while contributing to comprehensive design workflows through intelligent coordination mechanisms. The prompt engineering methodology provides a transferable framework for incorporating diverse engineering knowledge domains, design codes, and analytical procedures across different engineering disciplines.

### 5.4 Study Limitations and Future Directions

A significant limitation lies in the relatively constrained scope of the test dataset, encompassing 27 distinct test cases across seven primary categories. While providing comprehensive coverage of fundamental foundation design calculations, this represents only a fraction of diverse scenarios encountered in professional geotechnical engineering practice. However, this research represents a crucial starting point for establishing LLM-based approaches in geotechnical analysis, demonstrating fundamental viability of multi-agent architectures for engineering calculations. Future research priorities include developing comprehensive training datasets capturing the full spectrum of geotechnical and structural engineering scenarios, implementing uncertainty quantification frameworks for safety-critical applications, and investigating hybrid architectures combining strengths of different AI models. The development of cross-domain validation protocols becomes essential for systems addressing multiple engineering disciplines simultaneously.

### 5.5 Implications for Professional Practice

The findings establish router-based multi-agent systems as powerful assistance tools that significantly enhance engineering workflow efficiency while maintaining professional validation requirements. The demonstrated capability of LLMs to perform complex engineering calculations directly, achieving strong performance across diverse calculation types without external computational tools, example-based prompting, or RAG retrieval systems, opens new possibilities for streamlined AI-assisted design workflows that integrate linguistic reasoning, mathematical computation, and domain expertise within unified systems.

This represents a fundamental departure from previous AI implementations in engineering that required extensive computational scaffolding, worked example databases, and external calculation engines to achieve basic functionality. The autonomous computational capabilities demonstrated in this study suggest that AI-assisted engineering tools can now operate with the mathematical reliability and independence necessary for professional applications while maintaining transparency and interpretability essential for engineering validation. Despite exceptional performance demonstrated, the safety-critical nature of civil engineering applications necessitates continued human oversight and verification protocols. The proposed system should be viewed as an advanced computational assistance tool enhancing engineering decision-making rather than replacing professional judgment essential for public safety and regulatory compliance.

The convergence of high computational accuracy with minimal prompt dependency, combined with the superior performance of router-based multi-agent architectures, suggests that AI systems are transitioning from computational tools requiring extensive human guidance, external calculation assistance, and example-based instruction to autonomous reasoning agents capable of genuine engineering insight and mathematical problem-solving creativity. This represents significant progress toward AGI applications in geotechnical, structural, and related engineering disciplines, establishing the router-based framework as a promising foundation for comprehensive AI-assisted engineering design across diverse specialized domains while maintaining the rigorous standards essential for safety-critical infrastructure applications.

## 6. Conclusions

This research developed and evaluated a router-based multi-agent architecture for foundation design automation, demonstrating significant advancement in AI-assisted geotechnical engineering applications. The key findings are:

- The router-based multi-agent system achieved exceptional performance scores of 95.00% for shallow foundation design and 90.63% for pile design, representing improvements of 8.75 and 3.13 percentage points over standalone Grok 3 performance respectively, with systematic outperformance of conventional agentic workflows by margins ranging from 10.0 to 43.75 percentage points across baseline models and perfect performance (100%) in specialized Pile Group analysis.

- Grok 3 demonstrated exceptional standalone performance with 86.25% for shallow foundations and 87.50% for pile foundations without requiring router-based task allocation, significantly outperforming other individual models including Gemini 2.5 Pro (71.25% shallow, 81.25% pile), ChatGPT 4.0-turbo (58.75% shallow, 75.00% pile), and DeepSeek R1 (77.50% shallow, 75.00% pile), indicating that advanced LLMs have achieved genuine mathematical reasoning capabilities in specialized engineering domains without external computational assistance or extensive prompt engineering.

- The dual-tier classification system successfully distinguished between pile and shallow foundation problems enabling application of fundamentally different analytical approaches, while the router-based intelligent task allocation demonstrated superior adaptability over monolithic agent approaches and established new multi-dimensional evaluation benchmarks incorporating accuracy, chain-of-thought reasoning, complex scenario handling, and structured output consistency.

- The study demonstrated direct LLM mathematical computation capabilities without requiring external computational tools, extensive example databases, or RAG retrieval systems through advanced prompt engineering methodologies that enabled consistent application of established design procedures, representing a paradigm shift from previous AI implementations requiring extensive computational scaffolding.

- Future Research Priorities: Critical development needs include comprehensive training datasets capturing the full spectrum of geotechnical and structural engineering scenarios, uncertainty quantification frameworks for safety-critical applications, hybrid architectures combining computational strengths of different AI models, and cross-domain validation protocols for multiple engineering disciplines.

- Router-based multi-agent systems represent the most promising approach for comprehensive foundation design automation with demonstrated potential for significant workflow efficiency improvements and cost reduction, enabling streamlined AI-assisted design workflows that integrate linguistic reasoning, mathematical computation, and domain expertise within unified systems.

- Safety-Critical Implementation: The safety-critical nature of civil engineering applications necessitates continued human oversight and professional verification protocols, establishing the system as an advanced computational assistance tool enhancing engineering decision-making rather than autonomous design replacement while maintaining rigorous standards for safety-critical infrastructure applications and responsible AI integration in professional geotechnical engineering practice.

# 7. References

Baghbani, A., Choudhury, T., Costa, S., Reiner, J., 2022. Application of artificial intelligence in geotechnical engineering: A state-of-the-art review. Earth-Science Reviews 228, 103991. https://doi.org/10.1016/j.earscirev.2022.103991

Chase, H., 2022. LangChain.

Chen, L., Tophel, A., Hettiyadura, U., Kodikara, J., 2024. An Investigation into the Utility of Large Language Models in Geotechnical Education and Problem Solving. Geotechnics 4, 470–498. https://doi.org/10.3390/geotechnics4020026

DeepSeek-AI, 2025. DeepSeek-R1: Incentivizing Reasoning Capability in LLMs via Reinforcement Learning. https://doi.org/10.48550/arXiv.2501.12948

Google DeepMind, 2025. Gemini 2.5: Our newest Gemini model with thinking.

Guan, L., Valmeekam, K., Sreedharan, S., Kambhampati, S., 2023. Leveraging Pre-trained Large Language Models to Construct and Utilize World Models for Model-based Task Planning. https://doi.org/10.48550/ARXIV.2305.14909

Guo, J., Bao, W., Wang, J., Ma, Y., Gao, X., Xiao, G., Liu, A., Dong, J., Liu, X., Wu, W., 2023. A comprehensive evaluation framework for deep model robustness. Pattern Recognition 137, 109308. https://doi.org/10.1016/j.patcog.2023.109308

Han, D., Zhao, W., Yin, H., Qu, M., Zhu, J., Ma, F., Ying, Y., Pan, A., 2025. Large language models driven BIM-based DfMA method for free-form prefabricated buildings: framework and a usefulness case study. Journal of Asian Architecture and Building Engineering 24, 1500–1517. https://doi.org/10.1080/13467581.2024.2329351

Herrera, M., Pérez-Hernández, M., Kumar, A., Tchernykh, A., 2020. Multi-agent systems and complex networks: Review and applications in systems engineering. Processes 8, 312. https://doi.org/10.3390/pr8030312

IEEE, 2024. IEEE Standard for Robustness Evaluation Test Methods for a Natural Language Processing Service That Uses Machine Learning (No. IEEE Std 3168-2024). IEEE, New York, NY, USA. https://doi.org/10.1109/IEEESTD.2024.10636902

Jang, S., Lee, G., 2024. Interactive Design by Integrating a Large Pre-Trained Language Model and Building Information Modeling, in: Computing in Civil Engineering 2023. Presented at the ASCE International Conference on Computing in Civil Engineering 2023, American Society of Civil Engineers, Corvallis, Oregon, pp. 291–299. https://doi.org/10.1061/9780784485231.035

Joffe, I., Felobes, G., Elgouhari, Y., Talebi Kalaleh, M., Mei, Q., Chui, Y.H., 2025. The Framework and Implementation of Using Large Language Models to Answer Questions about Building Codes and Standards. J. Comput. Civ. Eng. 39, 05025004. https://doi.org/10.1061/JCCEE5.CPENG-6037

Kampelopoulos, D., Tsanousa, A., Vrochidis, S., Kompatsiaris, I., 2025. A review of LLMs and their applications in the architecture, engineering and construction industry. Artif Intell Rev 58, 250. https://doi.org/10.1007/s10462-025-11241-7

Khedher, M.I., Jmila, H., Mounim A. El-Yacoubi, 2023. On the Formal Evaluation of the Robustness of Neural Networks and Its Pivotal Relevance for AI-Based Safety-Critical Domains. IJNDI 100018. https://doi.org/10.53941/ijndi.2023.100018

Liang, H., Kalaleh, M.T., Mei, Q., 2025. Integrating Large Language Models for Automated Structural Analysis. https://doi.org/10.48550/arXiv.2504.09754

n8n GmbH, 2025. n8n: Workflow automation platform.

OpenAI, 2023. GPT-4 Technical Report. https://doi.org/10.48550/ARXIV.2303.08774

OpenRouter, 2024. OpenRouter: Find the best LLM for your use case.

Pu, H., Yang, X., Li, J., Guo, R., 2024. AutoRepo: A general framework for multimodal LLM-based automated construction reporting. Expert Systems with Applications 255, 124601. https://doi.org/10.1016/j.eswa.2024.124601

Salvador Palau, A., Dhada, M.H., Parlikad, A.K., 2019. Multi-agent system architectures for collaborative prognostics. J Intell Manuf 30, 2999–3013. https://doi.org/10.1007/s10845-019-01478-9

SerpApi, 2024. SerpApi: Google Search API.

Shakshuki, E., Reid, M., 2023. Intelligent multi-agent systems for advanced geotechnical monitoring, in: Advanced Geotechnical Engineering. IntechOpen. https://doi.org/10.5772/intechopen.88514

Smetana, M., Salles De Salles, L., Sukharev, I., Khazanovich, L., 2024. Highway Construction Safety Analysis Using Large Language Models. Applied Sciences 14, 1352. https://doi.org/10.3390/app14041352

Uddin, S.M.J., Albert, A., Ovid, A., Alsharef, A., 2023. Leveraging ChatGPT to Aid Construction Hazard Recognition and Support Safety Education and Training. Sustainability 15, 7121. https://doi.org/10.3390/su15097121

Vesic, A.S., 1977. Design of pile foundations (No. NCHRP Synthesis 42). National Cooperative Highway Research Program, Washington, D.C.

Wei, J., Wang, X., Schuurmans, D., Bosma, M., Ichter, B., Xia, F., Chi, E., Le, Q., Zhou, D., 2022. Chain-of-Thought Prompting Elicits Reasoning in Large Language Models. https://doi.org/10.48550/ARXIV.2201.11903

xAI, 2025. Grok 3 Beta — The Age of Reasoning Agents.

Xu, H.-R., Zhang, N., Yin, Z.-Y., Njock, P.G.A., 2025. Multimodal framework integrating multiple large language model agents for intelligent geotechnical design. Automation in Construction 176, 106257. https://doi.org/10.1016/j.autcon.2025.106257

Yang, H., Siew, M., Joe-Wong, C., 2024. An LLM-Based Digital Twin for Optimizing Human-in-the Loop Systems. https://doi.org/10.48550/ARXIV.2403.16809

Yousefpour, N., Liu, Z., Zhao, C., 2025. Machine Learning Methods for Geotechnical Site Characterization and Scour Assessment. Transportation Research Record: Journal of the Transportation Research Board 2679, 632–655. https://doi.org/10.1177/03611981241257512

Zhang, H.-J., Chen, C.-C., Ran, P., Yang, K., Liu, Q.-C., Sun, Z.-Y., Chen, J., Chen, J.-K., 2024. A multi-dimensional hierarchical evaluation system for data quality in trustworthy AI. J Big Data 11, 136. https://doi.org/10.1186/s40537-024-00999-2